\documentclass{optica-article}

\journal{opticajournal} % for journals or Optica Open
\usepackage{xr}
\articletype{Research Article}

\usepackage{lineno}
\usepackage{amssymb}
\usepackage{siunitx}
\usepackage[capitalize]{cleveref}
\crefname{section}{Sec.}{Secs.}
\Crefname{section}{Section}{Sections}
\Crefname{table}{Table}{Tables}
\crefname{table}{Tab.}{Tabs.}

\usepackage{float}
\usepackage{graphicx}
\usepackage{xspace}
\usepackage{textcomp}

\makeatletter
\DeclareRobustCommand\onedot{\futurelet\@let@token\@onedot}

\newcommand{\vect}[1]{\mathbf{#1}}

\def\de{\mathrm{d}}
\def\@onedot{\ifx\@let@token.\else.\null\fi\xspace}
\def\eg{\emph{e.g}\onedot} 
\def\ie{\emph{i.e}\onedot}

\def\bv{\mathbf{v}} 
\def\up{^{\prime}} 
\makeatother
\usepackage[OT2,T1]{fontenc}
\DeclareSymbolFont{cyrletters}{OT2}{wncyr}{m}{n}
\DeclareMathSymbol{\Sha}{\mathalpha}{cyrletters}{"58}

\makeatletter
\newcommand*{\rom}[1]
{\expandafter\@slowromancap\romannumeral #1@}
\makeatother
\begin{document}
\title{Neuromorphic Shack-Hartmann wave normal sensing}

\author{Chutian Wang,  Shuo Zhu, Pei Zhang, Jianqing Huang, Kaiqiang Wang and Edmund Y. Lam\textsuperscript{*}}

\address{Department of Electrical and Electronic Engineering, The University of Hong Kong, Pokfulam, Hong Kong SAR, China}

\email{\authormark{*}elam@eee.hku.hk} %% email address is required; see note below about the corresponding author designation

% use {asbstract*} to suppress the copyright line. Copyright information will be added in production
\begin{abstract*}\
The Shack-Hartmann wavefront sensor is widely employed in adaptive optics systems to measure optical aberrations. However, simultaneously achieving high sensitivity and large dynamic range is still challenging, limiting the performance of diagnosing fast-changing turbulence. To overcome this limitation, we propose neuromorphic Shack-Hartmann wave normal sensing (NeuroSH). NeuroSH is a unifying framework that harnesses the computational neuromorphic imaging paradigm to extract the high-dimensional wave normal from temporal diversity measurements. Both numerical analysis and experimental verification demonstrate the feasibility of NeuroSH. To the best of our knowledge, the proposed NeuroSH is the first scheme to surpass the optical dynamic range limitation under challenging dynamic scenarios, thereby advancing ultra-fast turbulence mitigation technology for cutting-edge imagers.
\end{abstract*}

%%%%%%%%%%%%%%%%%%%%%%%%%%  body  %%%%%%%%%%%%%%%%%%%%%%%%%%
\section{INTRODUCTION}
Adaptive optics (AO) is a technology extensively applied in various fields, including astronomy~\cite{roddier1999adaptive, hippler2019adaptive} and microscopy~\cite{booth2014adaptive, ji2017adaptive, gong2017optical}. Its primary objective is to enhance the performance of high-contrast imagers by integrating a wavefront sensor (WFS) for wavefront detection with a deformable mirror for wavefront compensation. Among the available WFSs, the Shack-Hartmann (SH) WFS stands out as the most widely adopted owing to its robustness and high optical efficiency. A SH WFS utilizes a microlens array (MLA) to sample the wavefront gradient as a combination of spots on the sensor plane, with their displacement linear to the corresponding local gradient. Hence, by comparing the registered reference position of SH spots and the aberrated one in the updated measurement, the map of the wavefront gradient can be recovered~\cite{platt2001history}.

Due to the fixed physical parameters in the utilized MLA and the pixelated sensor, the maximum and minimum magnitudes of the measured local tip/tilt term in the SH WFS are coupled~\cite{platt2001history}. The maximum value corresponds to the WFS dynamic range (DR), representing the range of detectable wavefront aberrations. While the minimum reflects the resolution of the sensor to detect subtle variations in the wavefront. This trade-off imposes significant constraints on the current design specifications, as achieving either high sensitivity or large DR becomes a compromise depending on the desired level and scale of aberrations of interest. However, current advanced AO systems pose a significant challenge if they necessitate the simultaneous achievement of large DR and high sensitivity. For instance, extreme AO systems aim to tackle fast-changing, large-scale, highly aberrated atmospheric turbulence under photon-starved and adverse conditions~\cite{fusco2016saxo, guyon2018extreme, beuzit2019sphere}. Consequently, extending the DR while preserving its sensitivity is a highly sought-after endeavor in the field.

Numerous DR enlargement methods have been proposed to enhance wavefront sensing, which can be broadly categorized into two groups: hardware-based and algorithm-based. The former requires considerable hardware modifications to the utilized MLA mask~\cite{saita2015holographic, shinto2016shack, kwon2020single, yi2021angle, lisingle,wu2024multiplexed}, leading to configuration bulkiness, manufacturing complexity, or higher expenses. Alternatively, it involves additional lateral or axial scanning strategies at the cost of sacrificing a substantial amount of temporal resolution~\cite{yoon2006large, aftab2018adaptive}. On the other hand, algorithm-based methods, such as sorting~\cite{lee2005sorting}, unwrapping~\cite{pfund1998dynamic}, Fourier modulation~\cite{carmon2003phase}, optical flow-tracking~\cite{vargas2014shack}, and deep learning~\cite{he2021deep, wang2024use}, offer advantages including cost-effectiveness and plug-and-play compatibility, without requiring additional modifications to its classical setup. However, their principle primarily revolves around avoiding the overlap of SH spots. This strategy imposes inherent restrictions on their DR enlargement capability, confining it within the limits of the optical dynamic range dictated by the wavefront curvature~\cite{akondi2021shack}.

To avoid the inherent ambiguity in the optical dynamic range without sacrificing temporal resolution, one prospective strategy is to leverage previous time-series shots. However, the conventional SH WFS employs an active pixel sensor (APS) that records the entire scene through a synchronous response mechanism, accompanied by large uninformative portions. The perceptual and statistical redundancies associated with the frame-based paradigm would require a significant waste of hardware storage and computational resources. In contrast, the emerging computational neuromorphic imaging (CNI) paradigm~\cite{pw_CNI2024} utilizes a bio-mimic dynamic vision sensor (DVS) as the data acquisition terminal. The DVS asynchronously responds to logarithmic intensity changes rather than capturing the complete scene~\cite{curtis2022event,zhang2024joint}. It offers microsecond temporal resolution and a high dynamic range exceeding~\SI{120}{dB}~\cite{brandli2014240}, which can potentially enhance the SH WFS performance~\cite{kong2020shack, wang2023tracking}.

In this paper, we propose neuromorphic Shack-Hartmann wave normal sensing (NeuroSH), which is a unifying CNI-based wavefront framework for directly perceiving the high-dimensional wave normal. NeuroSH is the first to surpass the current optical dynamic range limitation in a classical SH optical setup. We emphasize the advantage of CNI in terms of selective responses and low data redundancy, thanks to its asynchronous triggering mechanism. This enables the acquisition of a highly efficient data modality that represents the temporal diversity measurements within a complete AO servo-loop period. NeuroSH is therefore designed to perceive the spatio-temporal trajectory that approaches the temporal increment of wave normal.  The verification of NeuroSH is established on an adaptable SH setup for flexible MLA generation and dynamic phase ground truth overlaying. We further demonstrate its application in instantaneous flame front diagnosis, showcasing its substantial advantage over the DR limitations in ultra-fast scenarios.

\section{PRINCIPLE}

\subsection{SH WFS Geometry and the DR limitation}\label{sec: SH WFS}

\begin{figure}[ht!]
\centering\includegraphics[width=\textwidth]{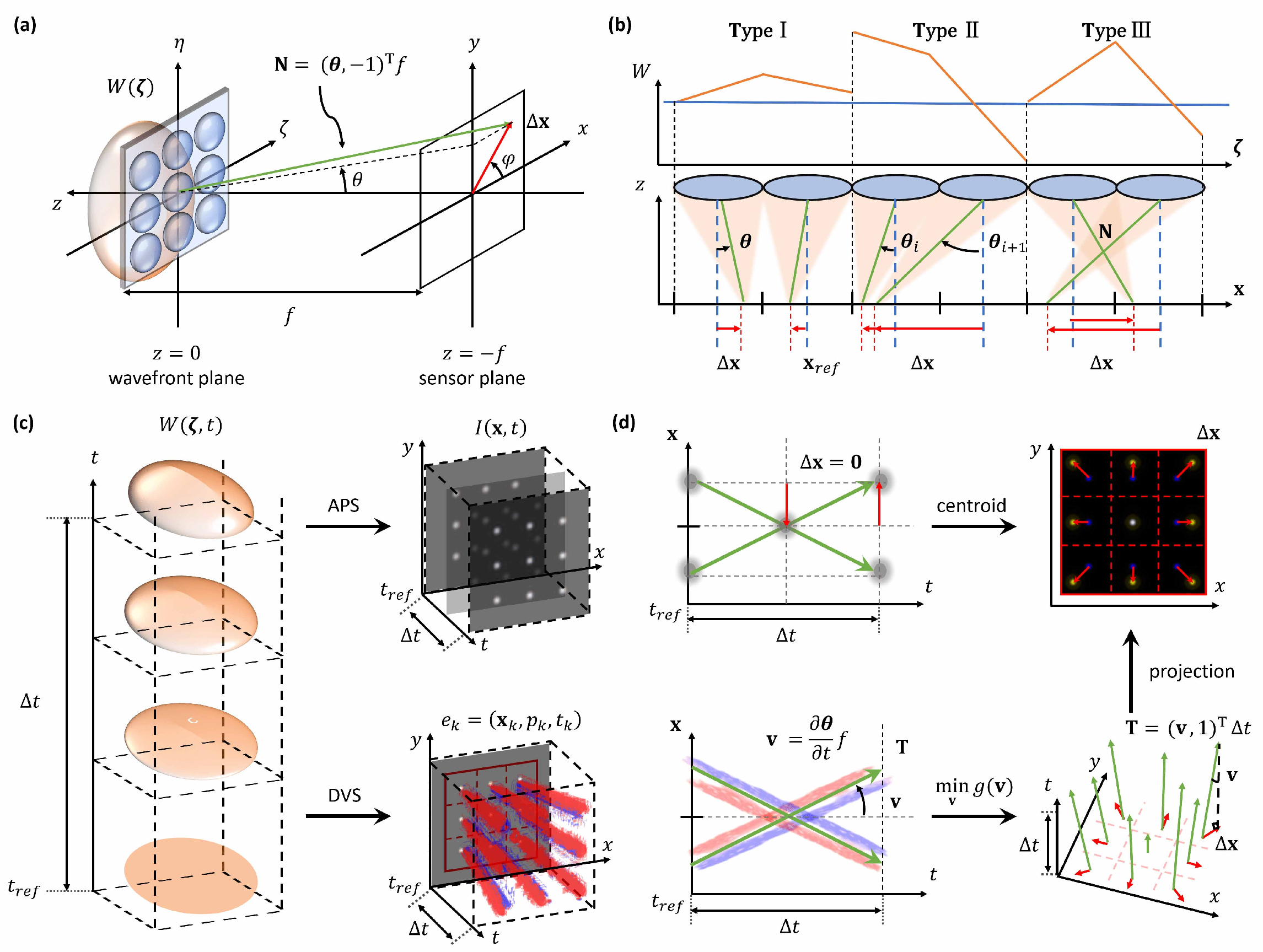}
 \caption{Principle of NeuroSH. (a) A geometric illustration of a classical SH WFS, where $\vect{N}$ is marked in green, and the displacement $\Delta \vect{x}$ is marked in red. (b) The DR taxonomy and the corresponding wavefront local shape (orange) compared with the reference (blue), The projection region of the lenslet boundary is marked with the major divisions on $\vect{x}$-axis. (c) NeuroSH gains dimensionality by leveraging the long-term information of a continuously changing wavefront, denoted as $W(\boldsymbol{\zeta}, t)$. In this temporal diversity strategy, utilizing an APS for acquiring frame stacks leads to significant redundancy, while the DVS asynchronously outputs a dense event stream with high efficiency, \ie, the NHG. (d) The conventional frame-based centroid tracking scheme is limited to measuring the 2D displacement $\Delta \vect{x}$ between adjacent frames due to the limited output channel capacity so that cannot handle Type \rom{2} and Type \rom{3} DR (top). In contrast, the NeuroSH approach directly estimates the spatio-temporal spot trajectory $\vect{T}$ in the NHG modality to approach the high-dimensional $\vect{N}$, overcoming the limitations of previous DR types. Upon solving for the optimal $\vect{T}$ through the minimization of the derived cost function $g(\vect{v})$ from the CNI paradigm, the 2D displacement $\Delta \vect{x}$ can be straightforwardly calculated as its projection (bottom).}\label{fig: NeuroSH}
\end{figure}

The wavefront is a surface of equal phase in Euclidean space $\mathbb{R}^3$. As depicted in Fig.~\ref{fig: NeuroSH}(a), considering the Cartesian coordinates of the wavefront plane as $\boldsymbol{\zeta} = (\zeta,\eta)$, we can express the distribution of the wavefront on this plane as a 2D function denoted by $W(\boldsymbol{\zeta})$. The wave normal, which is a 3D vector perpendicular to the wavefront, indicates the propagation direction of the light wave. Given $\theta$ and $\phi$ as the polar and azimuth angles for an incidence point, we can represent the wave normal as $(\boldsymbol{\theta}, -1)^{\mathsf{T}}$, where $\boldsymbol{\theta} = (\theta, \phi)^{\mathsf{T}}=\nabla W(\boldsymbol{\zeta})$ denotes a 2D angular incident vector.

A classical SH WFS comprises a MLA and a pixelated sensor placed at its focal plane, with its coordinates denoted by $\vect{x} = (x, y)$. The basic working principle of the SH WFS involves positioning the MLA at the desired wavefront plane to divide it into individual patches using the lenslet boundaries. In each patch, the regional wavefront is linearized to keep propagating along the direction of wave normal and is finally focalized to form a SH spot. The local principal ray vector is aligned with wave normal and can be therefore represented as $\vect{N} = (\boldsymbol{\theta}, -1)^{\mathsf{T}}f$, with the location of the SH spot being the end point of $\vect{N}$ on the sensor at the focal plane. Consequently, the linear relationship between the spot displacement and the angular incident vector can be written as
\begin{equation}\label{eq: SH WFS_geometry}
\Delta\vect{x}=\vect{x} - \vect{x}_{\textit{ref}}=\boldsymbol{\theta} f,
\end{equation}
where $f$ represents the focal length, and $\vect{x}_{\textit{ref}}$ serves as the registered position at initial stage. If assigning a plane wavefront as a reference, it corresponds to the lenslet's vertex within each patch.

As depicted in Fig.~\ref{fig: NeuroSH}(b), the fixed focal length and pitch size of the MLA used in SH WFS impose limitations on the detectability of the SH spots, referred to as the dynamic range (DR) of the SH WFS. Previously, two commonly adopted types of DR have been identified~\cite{platt2001history, carmon2003phase, akondi2021shack}, they are inherited in this work as Type \rom{1} and Type \rom{2} DR, respectively.  Herein, we further enlarge the DR to Type \rom{3} from the perspective of sensing the wave normal. The following  explicates each of the three types of DR in detail:

\begin{enumerate}
     \item Type \rom{1}. Spots can only be detected within their corresponding lenslet boundaries. As the detection of each spot is completely independent, it directly restricts the local maximum of the wavefront gradient, \ie,  $|\boldsymbol{\theta}|$.

     \item Type \rom{2}. Spots that stray outside the lenslet boundary can to some extent be detected based on avoiding their overlapping. This detection scheme is highly dependent on the association between adjacent spots. It restricts the local maximum of the wavefront curvature, \ie, the difference between $\boldsymbol{\theta}_{i +1}$ and $\boldsymbol{\theta}_{i}$.

   \item Type \rom{3}.  Spots can be detected even when their displacement paths overlap and cross over with each other in this work. The proposed scheme is motivated by recovering the high dimensional wave normal rather than its projection, thereby solving the ambiguity. The specific limitations are contingent upon the specific strategies employed to recover $\vect{N}$.
\end{enumerate}

\subsection{Strategies of Perceiving Wave Normal}\label{sec: WNS}
To perceive the high-dimensional $\mathbf{N}$, the fundamental step is to increase the dimensionality of the measurements. One commonly employed strategy is the implementation of axial scanning, which involves acquiring images from multiple planes along the $z$-axis. Several methods have been systematically investigated that leverage axial diversity measurements, including defocused SH WFS~\cite{li2014phase} and the known phase diversity~\cite{paxman1992joint,gonsalves2018phase}. Nevertheless, the introduced axial scanning scheme will inevitably sacrifice temporal resolution, significantly restricting its applicability in dynamic scenarios that require ultra-fast wavefront diagnosis.

Therefore, we develop a novel strategy to fully leverage the long-term information of the changing wavefront. It replaces the conventional active axial scanning scheme with a passive temporal-sequential measurement approach. To be specific, considering a temporally varying wavefront $W(\boldsymbol{\zeta}, t)$,  as shown in Fig.~\ref{fig: NeuroSH}(c). The changing rate of the wavefront will lead to the corresponding velocity of the spot for each lenslet, thus yielding a map of optical flow on the sensor plane
\begin{equation}\label{eq: SH2flow}
\begin{aligned}
    \nabla\big(\frac{\partial }{\partial t}W\big)&=\frac{\partial}{\partial t} \big(\nabla W \big) \\ 
    &=\frac{1}{f} \frac{\partial}{\partial t} \big(\boldsymbol{\theta} f\big) =\frac{1}{f}\frac{\partial}{\partial t} \big(\vect{x} -\vect{x}_{\textit{ref}} \big) \\
    &=\frac{1}{f}\frac{\partial \vect{x}}{\partial t} - 0  = \frac{\vect{v}(t)}{f},
\end{aligned}
\end{equation}
where the interchangeability of partial derivative order in the first line is established due to Schwarz’s theorem, and the equality in the second line holds according to~\Cref{eq: SH WFS_geometry}. 

Consequently, given two consecutive timestamps separated by an interval of $\Delta t$ (\ie, $t_1$ and $t_2$), the total increment of the angular incident vector $\boldsymbol{\theta}$ can be precisely reconstructed by directly integrating the instantaneous optical flow within $\Delta t$

\begin{equation}\label{eq: integ_t}
\frac{1}{f}\int_{t_1}^{t_2}  \mathbf{v}(t) \de t =  
\int_{t_1}^{t_2}  \frac{\partial\mathbf{\boldsymbol{\theta}}}{\partial t}\de t=\left. \boldsymbol{\theta}(t)\right |_{t_1}^{t_2}.
\end{equation}

If assigning $t_1$ as the reference timestamp, the measuring obtained by~\Cref{eq: integ_t} is simplified as an instantaneous estimation of its increment within $\Delta t$, \ie, $\Delta\boldsymbol{\theta}(t)$.

For a sufficiently small $\Delta t$ or uniform motion,~\Cref{eq: SH2flow} will trace a straight 3D motion trajectory $\vect{T}$ within a spatio-temporal neighbourhood, represented by 
\begin{equation}\label{eq: WNS}
\vect{T} = \Delta t (\bv , 1)^\mathsf{T} =\big((\frac{\partial\boldsymbol{\theta}}{\partial t}\Delta t)f,\Delta t\big)^\mathsf{T}\approx(\Delta \boldsymbol{\theta}f, \Delta t)^\mathsf{T},
\end{equation}
where the approaching on the right side is due to the finite difference approximation.  Note that~\Cref{eq: WNS} yields an expression similar to that of the scaled wave normal $\vect{N}$, \ie, $(\Delta \boldsymbol{\theta}f, -f)^\mathsf{T}$, with the third dimension now representing the temporal axis $t$ rather than the axial axis $z$. 

However, as shown in the top row of Fig.~\ref{fig: NeuroSH}(c), the classical WFS employs an active pixel sensor (APS) for data acquisition. The acquired temporal-varying Hartmanngrams form a stack of frames denoted as $I(\vect{x}, t)$, with the majority of values being non-informative zeros. Consequently, due to limited output channel capacity and computational memory, the significant redundancy in the frame modality inevitably restricts the maximum achievable sampling rate. Illustrated in the same row of Fig.~\ref{fig: NeuroSH}(d), the limited sampling rate diminishes the effectiveness of the temporal diversity scheme, reducing it back to a conventional 2D displacement tracking between adjacent frames through iterative looping. Consequently, the accurate reconstruction of the temporal derivative becomes challenging within the frame-based modality, ultimately leading to tracking failure in the case of Type \rom{2} and \rom{3} DR.

Hence, we propose leveraging the emerging CNI paradigm to facilitate a cheap temporal diversity strategy, \ie, NeuroSH. As depicted in the bottom row of Fig.~\ref{fig: NeuroSH}(c) and (d), NeuroSH employs a DVS to transform the redundant frame stack as a highly efficient neuromorphic data modality, termed the neuromorphic Hartmanngram (NHG). Thanks to the asynchronous triggering mechanism in CNI, the NHG benefits from inherent data sparsity and low redundancy. By approaching the temporal-varying scaled wave normal $\vect{N}$ as the trajectories $\vect{T}$ within a spatio-temporal neighborhood, NeuroSH is capable of perceiving the high-dimensional wavefront normal directly.

\subsection{Forward Model} \label{sec: model}
To derive a model for extracting the wave normal in utilizing the CNI paradigm, we begin by examining the fundamental event-triggering mechanism. As the DVS output, an event $e_{k} = (\vect{x}_{k}, p_{k}, t_{k})$ is triggered independently at the pixelated location $\vect{x}_{k} = (x_{k}, y_{k})$ if the logarithmic intensity change exceeds a predetermined threshold $C$ at the timestamp $t_k$~\cite{benosman2013event}. Herein, $k$ represents the counting index of each event point, and the symbol $p_{k} \in \{-1,+1\}$ indicates whether there is a positive ($p_{k}=+1$) or negative ($p_{k}=-1$) change. As such, the asynchronous event-triggering mechanism can be formulated as

\begin{equation}\label{eq: NI_trigger}
\Delta L(\vect{x}_{k}, t_{k}) \triangleq L(\vect{x}_{k},t_{k+1})- L(\vect{x}_{k}, t_{k})  = p_{k}C, 
\end{equation}
where $L(\vect{x}_{k}, t_{k}) \triangleq \log \big(I(\vect{x}_{k}, t_{k})\big)$ is defined as the logarithmic scale of the pixel-wise intensity $I(\vect{x}_{k}, t_{k})$, and $\Delta L(\vect{x}_{k}, t_{k})$ represents the recorded temporal increment from $t_k$ to $t_{k+1}$.

Without losing generality, considering one moving SH spot on the sensor, it will trigger a cluster of events on the pixels it traverses within $\Delta t$.  Such a motion-induced cluster is defined as $\mathcal{E}\triangleq\{e_{k}\}_{k=1}^{N_e}$, containing $N_e$ events.  Under static ambient illumination, the resulting temporal increment $\Delta L$ is completely caused by the moving of the spot pattern or the point spread function (PSF) of the sub-aperture. More details of theoretical derivation in Fourier Optics are included in Supplement 1. For a short $\Delta t$, the resulting event cluster $\mathcal{E}$ will enjoy a spatio-temporal correspondence due to the basic optical flow constraint~\cite{gallego2015event}. From that, we successfully link the measured increment to the physical prior of the reference target
\begin{equation}\label{eq: NI_flow}
\Delta L(\vect{x}) = -\nabla L(\vect{x}) \cdot \bv \Delta t,
\end{equation}
where $\Delta L(\vect{x})$ represents temporal increment in a short $\Delta t$, $\{\cdot\}$ indicates the dot product of two vectors, $\nabla L(\vect{x})=(\frac{\partial L}{\partial x},\frac{\partial L}{\partial y})^{\mathsf{T}}$ represents the spatial gradient of the target distribution near $t_\textit{ref}$.

In some naive approaches, the left-hand side of~\Cref{eq: NI_flow} is approximated by accumulating the individual increment contributed by each event to form an ``event frame'', \ie,
\begin{equation}\label{eq: NI_EF}
\Delta L(\vect{x}; \Delta t) = \int_{t_\textit{ref}}^{t_\textit{ref}+\Delta t} e_{k}(\vect{x}_{k}, p_{k}, t_{k}) \de t,
\end{equation}
where $e_{k}(\vect{x}_{k}, p_{k}, t_{k})=Cp_{k}\delta(\vect{x}-\vect{x}_{k})\delta(t-t_k)$ serves an alternative representative of an individual event that describes the amount of the carried increment.

\begin{figure}[ht!]
\centering\includegraphics[width=0.7\textwidth]{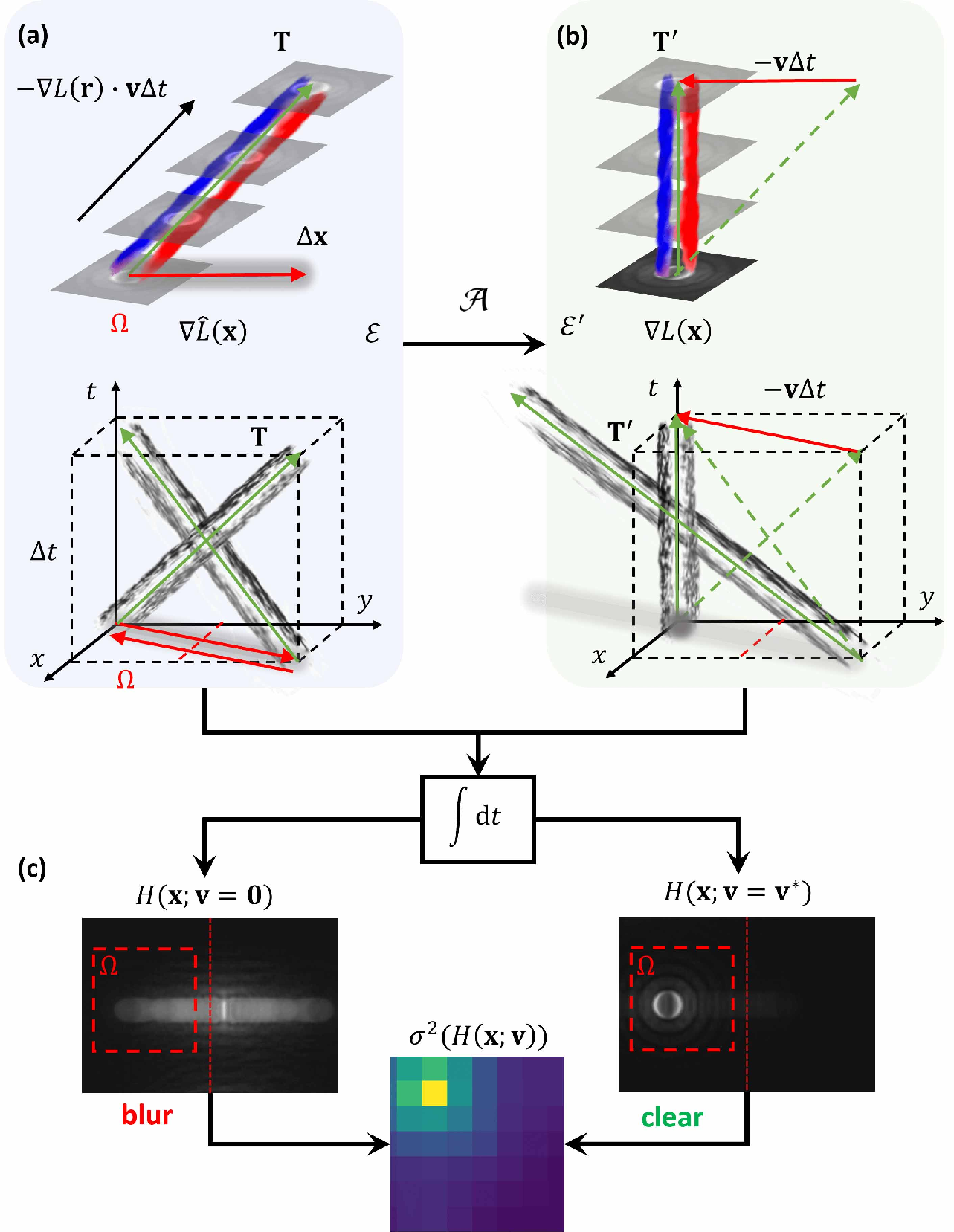}
 \caption{Algorithmic framework of NeuroSH. (a) The triggered SH event clusters $\mathcal{E}$. For each cluster, a simple embedding that maps the events to $\vect{T}$ can be determined by optimizing the warping parameter $\vect{v}$. The lenslet boundary $\Omega$ is indicated by red dashed lines. (b) The warped event cluster $\mathcal{E}\up$ is produced by applying an affine transformation $\mathcal{A}$. Here showcases the warping outcomes of an optimal candidate flow, where all the motion replicas are warped back to their reference positions. For (a) and (b), the top row depicts a single cluster (Type \rom{1}) while the bottom one represents the cases of cluster overlapping (Type \rom{2} and \rom{3}). (c) The contrast of IWE within its own $\Omega$ provides a metric to evaluate the alignment process in (b), which is robust to the cluster overlapping.}\label{fig: CMAX}
\end{figure}

However, the event frame is just a 2D image and will suffer from the loss of dimensionality. The direct accumulation along the $t$-axis will cancel the spatio-temporal correspondence and cause motion blur.  Therefore, we propose to manipulate the event clusters in NHG from a neural manifold perspective, as is discussed in~\cite{langdon2023unifying,munda2018real}.  Each event cluster is considered to describe a simple manifold embedding of the motion within the spatio-temporal neighborhood.  As shown in Fig.~\ref{fig: CMAX}(a), if consider a long period, the triggered event cluster $\mathcal{E}$ will contain replicas of the spatial gradient along the motion trajectory $\vect{T}$.   That is, the triggered events by the $i^{th}$ SH spots will ``grow'' along the motion trajectory $\vect{T}$ that starts from its own lenslet boundary $\Omega_{i}$ to form a distinct ``tubular'' morphology~\cite{benosman2013event}. As depicted in Fig.~\ref{fig: CMAX}(a), the tubular event clusters provide a distinction from each other in the whole NHG in the cases of Type \rom{1}, \rom{2}, and \rom{3} DR.

To extract the simple manifold embedding within NHG, a modified state-of-the-art contrast maximization framework is utilized to tackle the circumstance of cluster occlusion~\cite{gallego2018unifying,shiba2022secrets}. As is illustrated in Fig.~\ref{fig: CMAX}(b), the conventional contrast maximization applies an affine transformation $\mathcal{A}$ to the event clusters $\mathcal{E}$ in the NHG to produce a geometrically warped event cluster $\mathcal{E}\up\triangleq\{e_{k}\up\}_{k=1}^{N_e}$, with each point representing the warped event $e_k'$. The warping transformation can be defined as
\begin{equation}\label{eq: warp}
\vect{x}_k\up (\bv) = \mathcal{A}(\vect{x}_k; t_k, \bv) = \vect{x}_k - (t_k -t_{\textit{ref}}) \bv, 
\end{equation}
where the flow $\bv$ of the SH spot directly serves as the warping parameter.

Aggregating all the events in $\mathcal{E}\up$ yields an accumulation image, namely an image of warped events (IWE), \ie,
\begin{equation}\label{eq: NI_IWE}
H(\vect{x}; \bv) = \int_{t_\textit{ref}}^{t_\textit{ref}+\Delta t}e_{k}\up (\vect{x}\up_{k}, p_{k}, t_{k})\de t,
\end{equation}
where $e_{k}\up (\vect{x}\up_{k}, p_{k}, t_{k})=Cp_{k}\delta\big(\vect{x}-\vect{x}_{k}\up (\vect{v})\big)\delta(t-t_k)$ represents the discrete warped event by applying the affine transformation $\mathcal{A}$, with $\delta (\cdot)$ being the Kronecker delta function. 

By substituting~\Cref{eq: NI_trigger} and ~\Cref{eq: NI_flow} into~\Cref{eq: NI_IWE},  the IWE can be interpreted as
\begin{equation}\label{eq: NI_IWE v.s. flow}
H(\vect{x}; \bv) =-\nabla \hat{L}(\vect{x}) \cdot\bv\Delta t,
\end{equation}
where $\nabla \hat{L}(\vect{x})$ represents the estimated spatial gradient.

The contrast maximization assumes that if all the events triggered by the moving edges are properly aligned, the replica of these edges will be warped back to the reference state. This process is referred to as motion compensation, as all the moving edge replicas after warping now share the same pixel coordinates. Hence, as depicted in Fig.~\ref{fig: CMAX}(c), if $\bv =0$, then no motion is compensated, the producing IWE, \ie, $H(\vect{x}; \bv =0)$ is equivalent with the definition of event frame, according to~\Cref{eq: NI_EF}. This is a histogram that merely estimates a blurred gradient. On the other hand, if the candidate flow is the optimal one $\bv^{*}$, the yielded IWE will produce the clearest estimation of the PSF spatial gradient, \ie, $\nabla \hat{L}^*(\vect{x}) = \nabla L (\vect{x})$, given a specific $t_{\textit{ref}}$. The expression in~\Cref{{eq: NI_IWE v.s. flow}} can be therefore rewritten as
\begin{equation} \label{eq: blur_free}
    H(\vect{x}; \bv=\bv^*) = - \nabla L (\vect{x})\cdot \bv^{*}\Delta t,
\end{equation}
where $\nabla L (\vect{x})$ represents the motion-resolved spatial gradient.

This problem is thereby transformed into an optimization problem by selecting an appropriate cost function to evaluate the contrast. According to the performance comparison list offered in \cite{gallego2019focus}, the variance of IWE is the lowest time-consuming (only $\SI{16.9}{\mu s}$ ) and the most robust type of loss. The calculation of variance is also independent of the specific pattern, which indicates that it is a unifying metric. For each flow $\vect{v}$ starting from its own $\Omega$, the selected cost function can be defined individually 
\begin{equation}\label{eq: NI_cost}
g(\vect{v})=-\sigma^2\big(H (\vect{x};\bv)\big) = - \frac{1}{|\Omega|} \int_{\Omega} \bigg(H - \mu\big(H (\vect{x};\bv)\big)\bigg)^2  \de \vect{x},  
\end{equation}
where $|\Omega|$ is the area within the lenslet boundary, $\mu(\cdot)$ represents the statistical mean.

\Cref{eq: NI_cost} indicates that there is no need to pre-separate the NHG into different clusters $\mathcal{E}$. To be specific, for a flow allocated to the $i^{th}$ lenslet, denoted as $\vect{v}_i$, the sub-cost function $g(\vect{v}_i)$  can be computed by first applying a global warping transform to the entire NHG and then examining the cropped accumulated IWE in its own $\Omega_i$. This derives an overall evaluation by enumerating all $N_{\textit{MLA}}$ lenslets and summing their sub-costs. By incorporating it with~\Cref{eq: SH2flow}, we link the flow to the angular incident vector, thus formulating the forward model, \ie, 

\begin{align}\label{eq: opt}    
\begin{aligned}\mathop{\text{minimize}}_{\boldsymbol{\theta}_i,\vect{v}_i} & \quad \sum_{i = 1}^{N_{\textit{MLA}}} g(\boldsymbol{\theta}_i)\\
    \text{subject to} & \quad 
    f\frac{\partial \boldsymbol{\theta}_i }{\partial t}=\vect{v}_i.
    \end{aligned}
\end{align}

% + h(\vect{v}_1,\vect{v}_2,...,\vect{v}_{N_{\textit{MLA}}}) . $h(\cdot)$ represents some common regularizers on the distribution of the flow map, for example, the smoothness term $|\cdot|_2^2$ in ~\cite{horn1981determining}.
Consequently, since the model that links the physical optics to the neuromorphic modality is differentiable, the optimal solution can be easily extracted using well-established optimization algorithms in PyTorch~\cite{ kingma2014adam, wang2022differentiable}. To further accelerate the iterative optimization, other advanced convex optimization variants~\cite{boyd2011distributed,song2022dual} and learning-based algorithms can be integrated~\cite{zhang2024neuromorphic}.

\section{EXPERIMENT AND RESULTS}
\subsection{Experimental Verification}

\begin{figure}[t!]
\centering\includegraphics[width= 0.95
\textwidth]{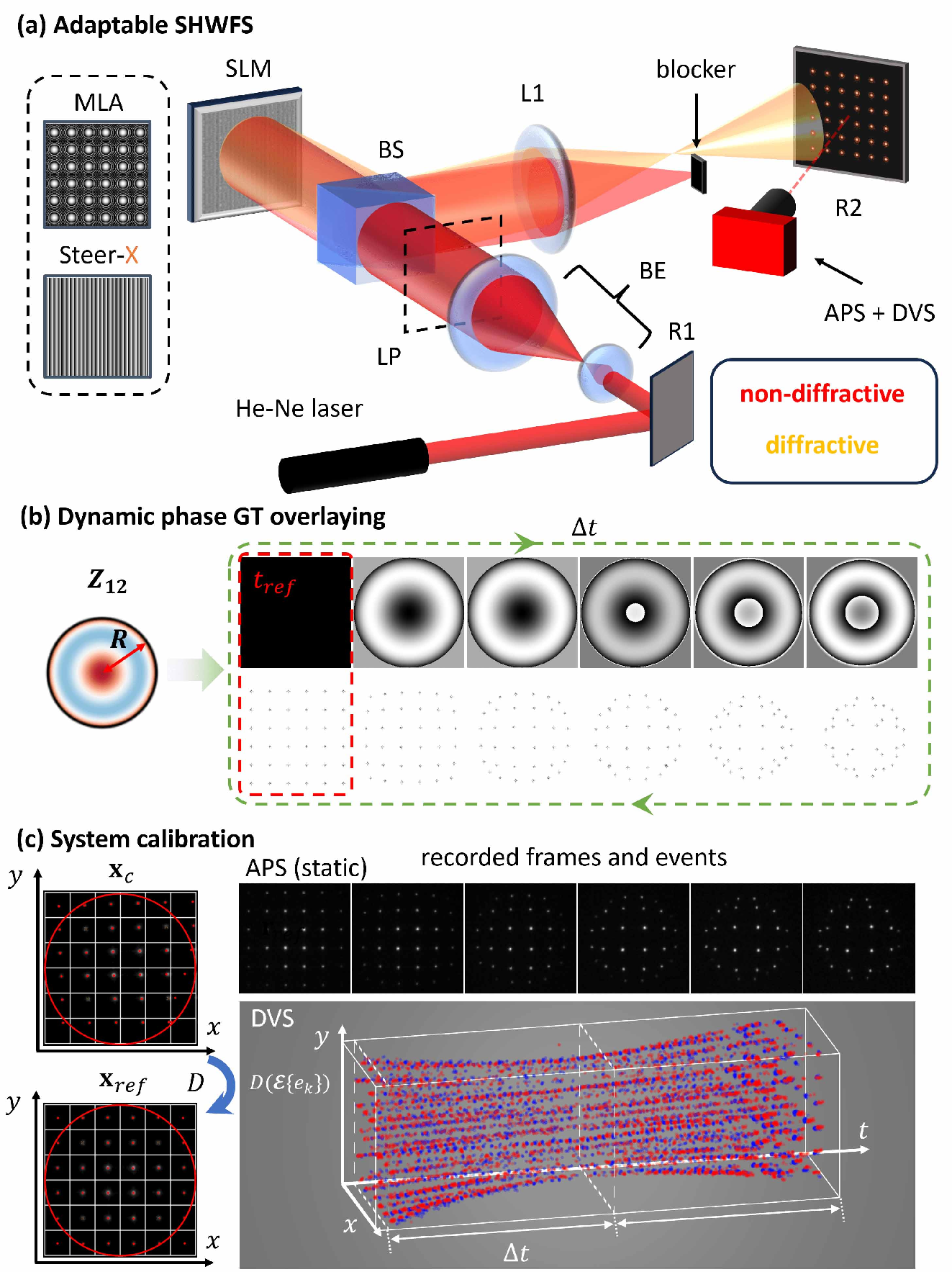}
      \caption{Experimental verification of NeuroSH implemented on an adaptable SH setup. (a) The optical configuration of the adaptable SH WFS. SLM: spatial light modulator, BG: blazed grating, BE: beam expander, LP: linear polarizer, BS: beam splitter, L1: positive lens, R1: mirror reflector, R2: diffusive reflector. (b) The dynamic loading of the phase GT. $R$: Zernike radius, whose conjugate area on the sensor plane is marked with a red ring in (c). (c) Outcome of system calibration. APS outputs the frame under long exposure and the DVS outputs dense events to form the NHG in dynamic measurement. $D$: calibration matrix.} 
\label{fig: adaptable_SH}   
\end{figure}

To verify the feasibility of NeuroSH, we conducted an experimental validation of NeuroSH using a self-built adaptable SH setup, as depicted in Fig.~\ref{fig: adaptable_SH}. The adaptable SH setup offers two significant advantages. The prominent advantage is the ability to generate a series of dynamic known patterns on the same plane as the MLA, thereby providing a reliable phase ground truth (GT) for the temporal diversity scheme. Additionally, the setup allows for flexible overlaying of the Fresnel diffractive lens array pattern on the spatial light modulator (SLM), enabling a high degree of freedom in arranging the simulated MLA configuration~\cite{lechner2020adaptable}.

As depicted in Fig.~\ref{fig: adaptable_SH}(a), the adaptable SH setup incorporates a reflective phase-only SLM (LETO-3, $1920 \times 1080$ pixels, $\SI{6.4}{\mu m}$) as the core component. In the setup, a He-Ne laser (Thorlabs, HNL100L, $\SI{632.8} {nm}$) is utilized for coherent illumination. The emitted laser beam passes through a $10 \times$ beam expander (Thorlabs, BE10M-A) and a linear polarizer to ensure collimated and linearly polarized illumination. To account for the spatial limitations imposed by the beam splitter used in the reflection arm, a relay lens L1 with a focal length of $\SI{15}{cm}$ is positioned to relay the Hartmanngram to an intermediate reflector (R2). Subsequently, another camera lens L2 is employed to further relay the light onto the sensor plane of an event camera (iniVation, DAVIS 346). The event camera has a resolution of $346 \times 260$ pixels and supports both APS and DVS output modes. 

In practice, the diffractive efficiency of the reflective SLM is typically less than \SI{67}{\%}, resulting in the formation of a significant amount of zeroth-order light from the dead areas of the SLM. To address this problem, a steer modulation technique is employed~\cite{zhang2009elimination}. This technique involves superposing a 2D linear phase pattern, also known as a blazed grating, onto the entire SLM screen. The added steer modulation aims to induce a lateral frequency shift on the Fourier plane, separating the diffractive reflective beam from the zeroth-order reflective beam (non-diffractive). To select the diffracting part of the reflective beam and exclude the higher-order terms caused by the periodic SLM pixels, a blocker is introduced on the Fourier plane. Additionally, due to the special logarithmic response curve characteristic of the event-triggering mechanism in CNI, the output from the DVS channel exhibits high sensitivity to ambient noise as well as the adjustment of the liquid crystals in the utilized SLM.  Therefore, an additional step is introduced to further relay the Hartamanngram to an intermediate diffusive reflector R2. This allows for flexible adjustment of background illumination, thus leveraging the optimal segment of the logarithmic curve.

The generation of the Fresnel diffractive lens array pattern onto the SLM follows the principle described in~\cite{zhao2006efficient}. In our setup, we generated a $6 \times 6$ Fresnel lens phase pattern with a focal length of \SI{10}{cm} and a pitch size of $180$ SLM pixels. This design allows us to make full use of the limited spatial resolution of the SLM. The system magnification is approximately 0.5, which resizes the SLM aperture with a size of \SI{6.9}{mm} onto our sensor with a size of less than \SI{4.8}{mm}. As a result, each MLA pitch corresponds to approximately 31 sensor pixels. While using a singlet to relay the Hartmanngram onto R2 introduces additional phase aberration, it is important to note that SH WFS measures the wavefront increment rather than its absolute value. Therefore, even if the position of $r_{\textit{ref}}$ slightly deviates from the center of each sub-aperture due to the introduced aberration, the linear relationship remains valid.

In terms of overlaying the dynamic phase GT, the $12^{th}$ order Zernike polynomial, \ie, the spherical aberration $Z_{12}$, is chosen to simultaneously evaluate all three types of DR. As shown in Fig.~\ref{fig: adaptable_SH}(b), the normalized phase distribution of $Z_{12}$ is characterized in polar coordinates within a Zernike radius of $R$. We then generate a series of patterns corresponding to $Z_{12}$, with their amplitude increasing linearly in the range of $[0, 5] \pi$ rad. To achieve optimal continuity, we ensure a maximum SLM refreshing rate of $\SI{60}{Hz}$ and adjust the step to $0.2 \pi$ rad. This results in a loop period $\Delta t$, which is nearly \SI{400}{ms}. In the top right of Fig.~\ref{fig: adaptable_SH}(b), we present a series of images corresponding to the $Z_{12}$ GT and its resultant simulated Hartmanngram within the same range but with a larger step of $\pi$ rad for clearer demonstration. With an increase in mode amplitude, the edge region of $Z_{12}$ represents a sharper change, particularly around the "valley" region. This leads to contrary increasing directions of the local spatial gradient, gradually being dominated by Type \rom{2} and \rom{3} DR. The system calibration of NeuroSH is illustrated in the top right of Fig.~\ref{fig: adaptable_SH}(c). The specific procedures involve identifying a calibration matrix, denoted as $D$, by fusing the long exposure frame from the APS channel under static conditions and applying it to the dynamically triggered event streams. More details are included in Supplement 1.

\subsection{Algorithmic Implementation}

\begin{figure}[ht!]
\centering\includegraphics[width=
\textwidth]{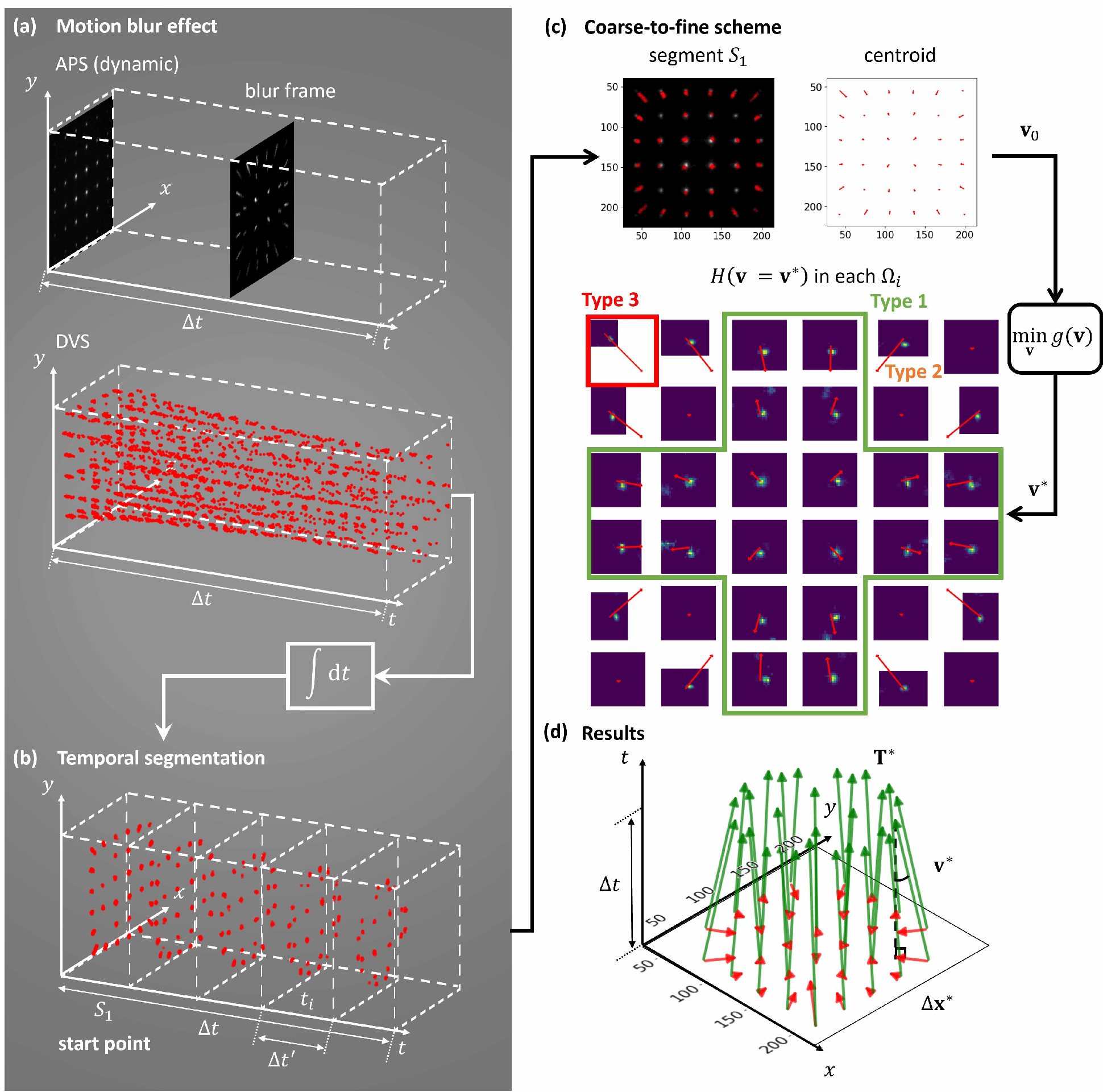}
\caption{Algorithmic implementation of NeuroSH. (a) A comparison of the APS and DVS output in dynamic scenarios. (b) An event frame $S_i$ is calculated within the $i^{th}$ temporal segment with an interval of $\Delta t'$. (c) The coarse-to-fine alignment scheme in the NeuroSH workflow. The NeuroSH takes the robust centroid of the first segment $S_1$ as the starting point for coarse alignment and performs finer tuning to the long-term events by utilizing the global warping scheme. Here shows the compensated IWE of each $\Omega$ according to $\vect{v}$. In the cases of Type \rom{2} and \rom{3} DR, the spot crosses the boundary.  For better results illustration, the fixed region of $\Omega$ (marked in dark blue) is shrunk by normalizing the maximum displacement vector. (d) Extraction outcomes of the optimal 3D trajectories $\vect{T}^*$ (green) and their projection on the sensor plane,  \ie, the 2D displacements $\Delta\vect{x}^*$ (red).}  
\label{fig: results1}   
\end{figure}

Fig.\ref{fig: results1}(a) illustrates the performance of the event camera from different channels in practical dynamic scenarios. The APS channel preserves an exposure time of \SI{300}{ms} to adapt to the ambient illumination, however, it suffers from significant motion blur. Therefore, after completing the calibration procedure and registering the reference position $\vect{x}_{\textit{ref}}$, only the DVS output is utilized in the algorithmic implementation step. Additionally, to ensure a fair performance comparison with\cite{kong2020shack}, only positive events, \ie, $p_k = +1$, are utilized.

The coarse alignment is realized by adopting conventional centroid methods through a temporal segmentation scheme, similar to that in~\cite{kong2020shack}. Specifically, for any timestamp $t_i$ of interest, the events within the spatio-temporal neighborhood are divided into several temporal segments using a smaller time window $\Delta t'$. The events within adjacent segments are then accumulated according to~\Cref{eq: NI_EF} to produce the segmented event frames $S_i$ and $S_{i+1}$. The displacement between the centroids of $S_i$ and $S_{i+1}$ within each $\Omega$ serves as the starting point. As depicted in Fig.~\ref{fig: results1}(b), for the reference timestamp, since the reference position has already been determined in the calibration step, only the centroid of $S_1$ needs to be calculated. A robust starting point, in addition to helping the optimizer avoid being trapped in a local minimum due to flow discontinuity, also aids in determining whether a reliable motion is occurring within each $\Omega$. For instance, in this case, we have identified 7 sub-apertures to be excluded in the following optimization, as they contain events that fall below a certain threshold in the first segment. Consequently, these points are filtered out from further analysis.

The fine alignment procedure involves applying the proposed NeuroSH forward model to perform iterative optimization.  As shown in the bottom of Fig.~\ref{fig: results1}(c), the combined figure contains the optimal candidate flow as well as the corresponding IWE. By examining the contrast of IWE, the effectiveness of NeuroSH has been demonstrated.  To showcase that NeuroSH has successfully tackled all three types of DR, we preserve an independent view of each extracted flow starting from each lenslet boundary. In the case of Type \rom{2} and \rom{3} DR, the fixed region of each $\Omega$, colored in dark blue, is scaled to adapt to the displacement outside the lenslet boundary. The independent views of Type \rom{1}, \rom{2}, \rom{3} DR are marked with green, orange, and red, respectively. Fig.~\ref{fig: results1}(d) illustrates the optimization outcome for all the reliable starting points. Herein, the extracted optimal trajectory $\vect{T}^*$ is straight within the long period $\Delta t$ due to the linear increasing of the $Z_{12}$ mode. It is worth noticing that this verification procedure is still fundamental as well as effective in non-linear scenarios through basic temporal segmentation.

\begin{figure}[ht!]
\centering\includegraphics[width= 0.6
\textwidth]{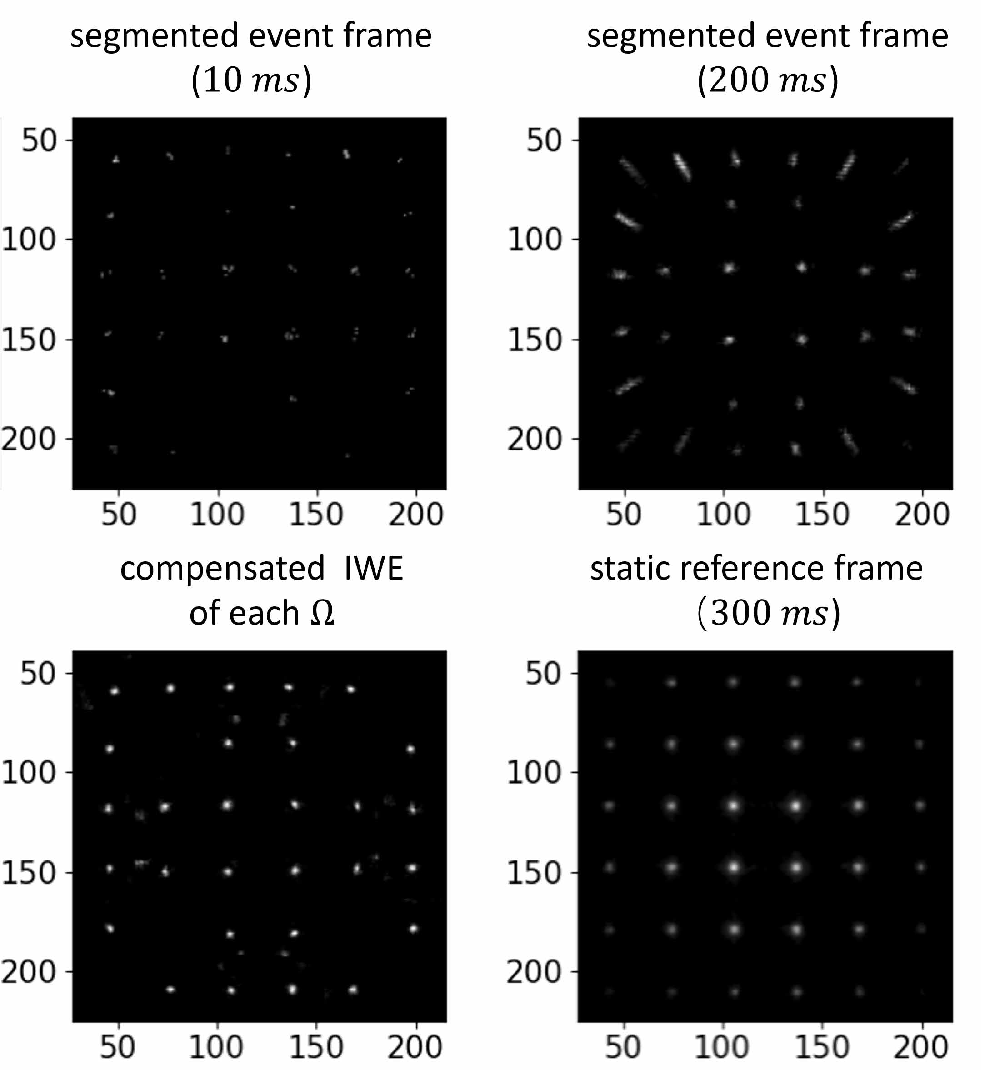}
\caption{Upper row: The produced event frames within the temporal segment of different sizes. Lower row: The combination of the compensated IWE of each $\Omega$ helps get rid of the motion blur and enhance the contrast under adverse ambient illumination, compared with the reference frame captured in the static scenario with an exposure time of nearly \SI{300}{ms}. All sub-figures have been normalized for the clearest demonstration.} 
\label{fig: rec_frame}   
\end{figure}

Additionally, NeuroSH exhibits further superiority in reconstructing the motion-resolved Hartmanngram with low-light enhancement compared with the conventional methods based on the event frame. As shown in Fig.~\ref{fig: rec_frame}, given a selected timestamp of interest, say $t_{\textit{ref}}$, the production of an event frame requires a manual selection of the segmentation window $\Delta t'$. This is prone to be problematic in general cases where both fast and slow flows occur simultaneously within one single segment. On the one hand, to catch the fast flow,  $\Delta t\up$ should be chosen as short as possible. However, this leads to a low signal-to-noise (SNR) in such a thin segment. For example, setting $\Delta t'$ as \SI{10}{ms} makes the spots in the center of the event frame nearly invisible, since each event naturally conveys less information and is subject to noise. On the other hand, choosing a larger window, \eg, setting $\Delta t\up$ to \SI{200}{ms}, will again result in the motion blur effect to the spots near the edge. The NeuroSH tackles this challenge by combining the compensated IWE of each $\Omega$ together to reconstruct a HDR image that is free from motion blur. The comparative outcome between the yielded IWE and the APS reference frame showcases the robustness of NeuroSH, owing to its weak reliance on the selection of $\Delta t$.

\subsection{Results Comparison and Error Analysis}
\begin{figure}[ht!]
\centering\includegraphics[width=0.62
\textwidth]{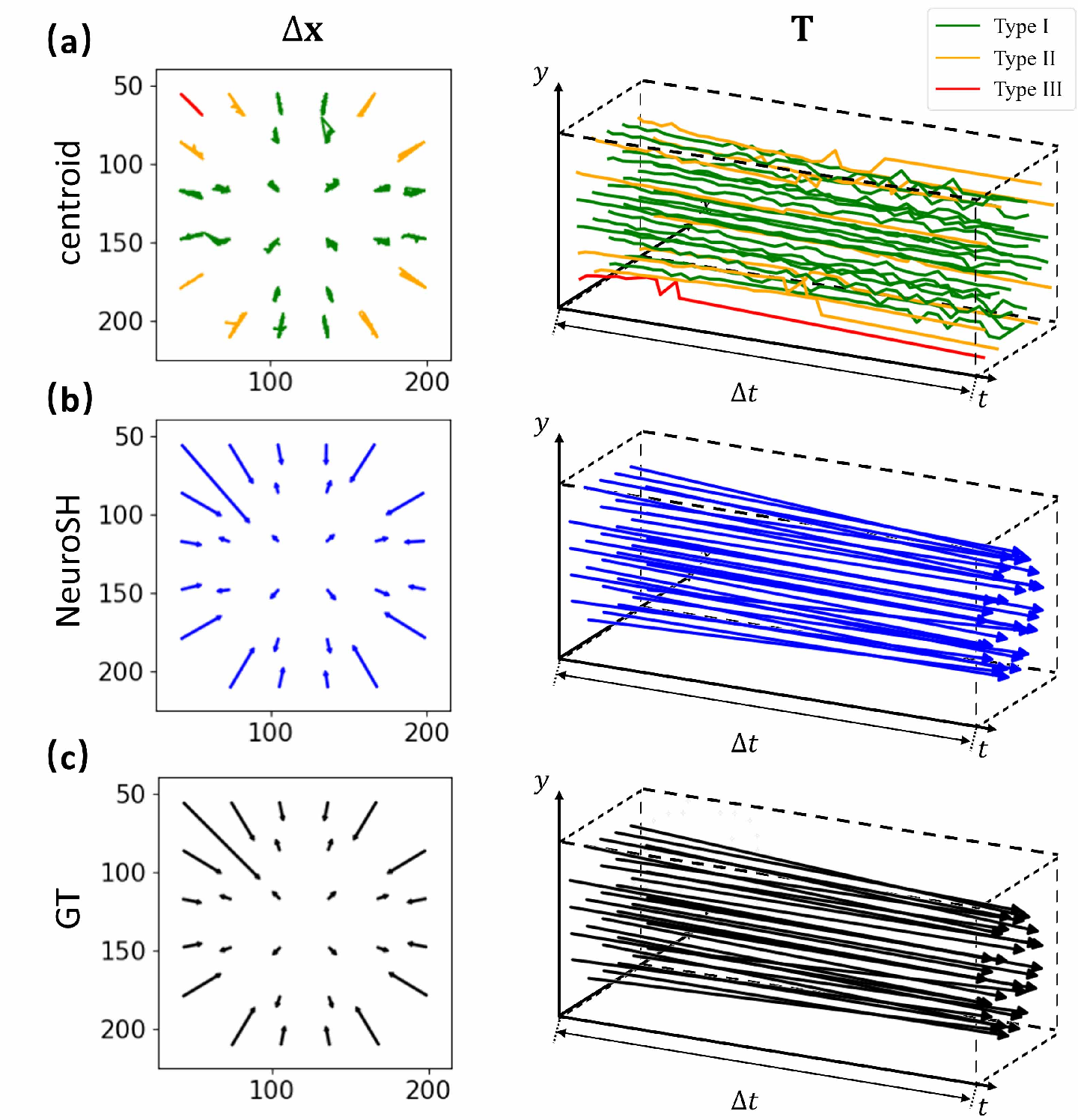}
\caption{Results of the extracted 2D velocity vectors and the 3D trajectories. (a) The outcome tracked by the classical centroid method with $\Delta t\up$ set as \SI{10}{ms}, where the Type \rom{1}, \rom{2}, \rom{3} DR marked in green, orange, red, respectively. (b) The NeuroSH outcome is marked in blue, illustrating that all three types of DR can be handled. (c) The calculated GT is marked in black.} 
\label{fig: tracking}   
\end{figure}

To analyze the performance of NeuroSH, we reproduce the conventional centroid tracking method to form the comparative group. Specifically, by accumulating events within a segmentation time window $\Delta t'$ of \SI{10}{ms} to form the event frame, the centroid in each $\Omega$ can be calculated. As shown in Fig.~\ref{fig: tracking}, the 3D trajectories of the SH spots can be extracted through this temporal tracking scheme. To highlight the centroid tracking outcomes in terms of different types of DR, we colored the trajectory of Type \rom{1}, \rom{2}, and \rom{3} DR in green, orange, and red, respectively. By comparing the outcomes of the NeuroSH with that of GT, it is evident that the proposed NeuroSH model unequivocally provides rapid and dependable tracking for all types of DR. Conversely, the results obtained from classical centroid tracking cannot deal with  \rom{2}, and \rom{3} DR and exhibit a notable fluctuation attributable to the low SNR within each thin segment and cross-talk from adjacent sub-apertures.

\begin{figure}[ht!]
\centering\includegraphics[width= 0.6
\textwidth]{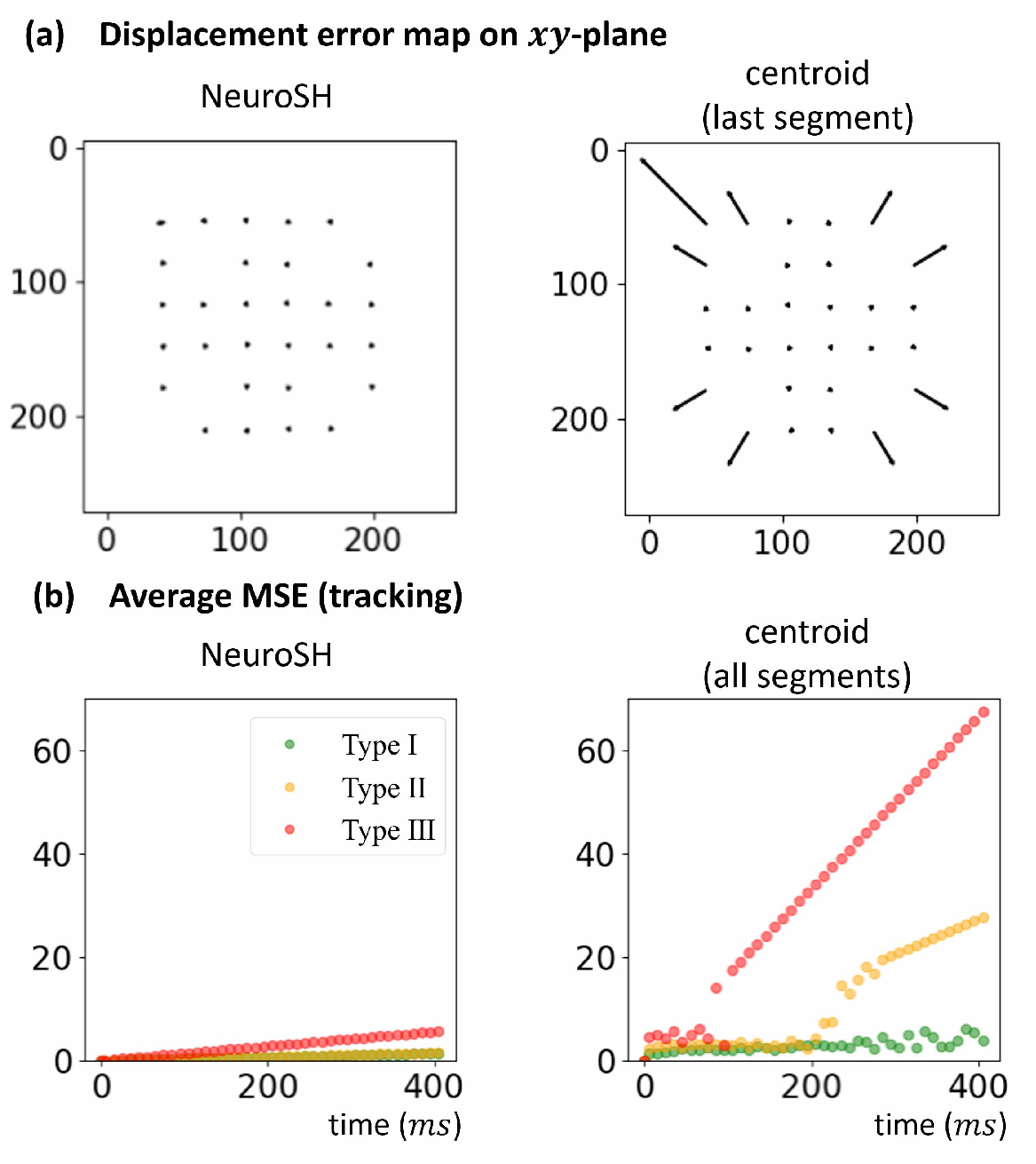}
\caption{Error analysis. (a) The endpoint 2D displacement error map on $xy$-plane with the outcome from NeuroSH (left) and the classical centroid method applied on the last segment. (b) The tracking mean squared errors in a full loop $\Delta t$. For each type of DR, the error value is averaged on all sub-apertures that are belonged to. Type \rom{1}, \rom{2}, \rom{3} are marked in green, orange and red dots, respectively.}
 
\label{fig: error}   
\end{figure}

In error analysis, we first examine the end-point measurement outcome, that is, to extract the displacement map directly from the last segment in Fig.~\ref{fig: error}(a). The centroid outcome calculated with a $\Delta t'$ of \SI{10}{ms} is extremely noisy. To guarantee a fair comparison,  the last segment window $\Delta t'$ is tuned to \SI{80}{ms} to exact the tracking of the spots as much as possible. However, as discussed before, the challenge of precisely determining the fast and slow flow simultaneously within the produced event frame is impossible. Furthermore, it is clear that nearly no displacement can be detected by the centroid method for spots allocated to Type \rom{2} and \rom{3} DR. Fig.~\ref{fig: error}(b) illustrates the mean squared errors on the average of the sub-apertures allocated to the 3 types of DR for all segments. It is obvious that once the points move outside their designated aperture $\Omega$, the centroid tracking fails and exhibits significant tracking errors.  Consequently, the advantages of NeuroSH in tracking spots demonstrate its capability of achieving accurate and robust wavefront sensing that surpasses the previous DR limitations.

\subsection{Application: Instantaneous Flame Front Visual Odometry}\label{sec: Flame front}

Here we exhibit an application of the proposed NeuroSH for instantaneous flame front visual odometry. Flame front visualization provides valuable insights into the physics and chemistry of flames, such as flame structure, flame stability, flame propagation, and reaction kinetics. A flame front is defined as the boundary or interface separating the unburned fuel from the combustion products within a propagating flame. It represents an active area where the combustion reaction takes place, exhibiting a rapid emission of energy, heat, and light. That is, its specific sub-region encompasses a significant temperature gradient, resulting in a large-gradient wavefront. As a result, the observation and analysis of the flame front's intricate structure during the dynamic propagation process necessitate the wavefront sensing technology with a large DR.

\begin{figure}[ht!]
\centering\includegraphics[width= 1
\textwidth]{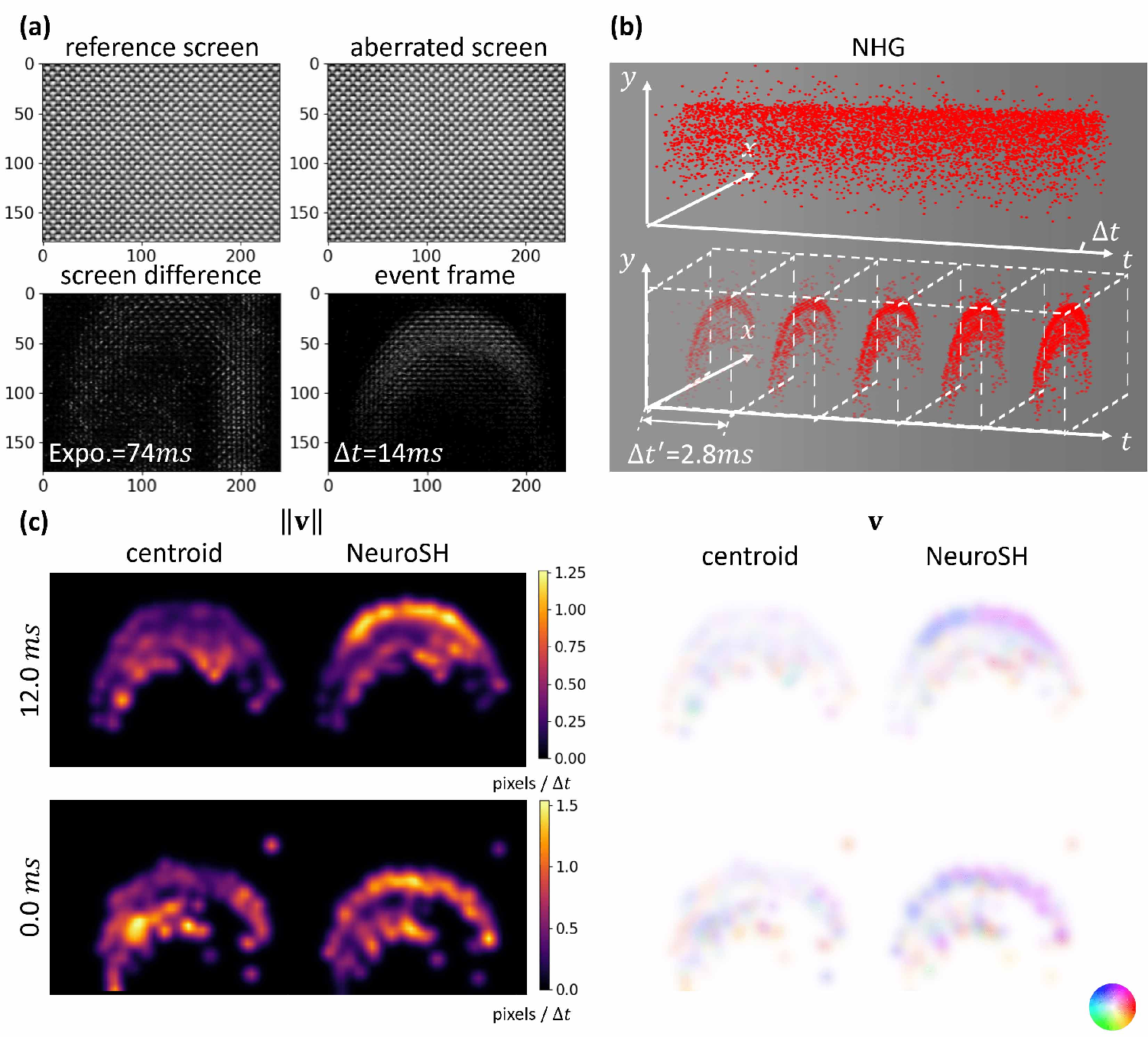}
\caption{Instantaneous flame front diagnosing in utilizing the proposed NeuroSH framework. (a)The output from APS and DVS channels and the accumulated event frame within $\Delta t$ for comparison. (b) The full NHG modality within $\Delta t$. (c) The extracted absolute velocity map and the directional pseudo-color velocity map using conventional centroid tracking and NeuroSH. The $36\times 48$ flow map has been smoothed by applying a Gaussian blur kernel.} 
\label{fig: results2}   
\end{figure}

To achieve flame front characterization, the $6\times6$ sampling grid used in the GT verification experiment is inadequate for finer sampling. Due to the limited resolution of the phase-only SLM, generating a higher-dimensional MLA is a great challenge. In addition, The flame front visualization should necessitate a ten times larger field of view, which cannot achieved for the current SLM screen without scaling. Referring to~\cite{bichal2013application, raffel2015background},  the background-oriented schlieren (BOS) technology shares a similar gradient-based forward model with the classical SH wavefront sensing, but it offers a larger FOV. We implemented the BOS experiment by generating a chessboard mask on the screen to simulate a densely hexagonally arranged MLA with a $36 \times 48$ sampling rate.  More implementation details about the BOS setup are contained in Supplement 1.

As illustrated in Fig.~\ref{fig: results2}(a), the APS channel's fixed exposure time of approximately \SI{74}{ms} restricts its ability to capture anything beyond blurred aberrated screens in dynamic scenarios. Hence, the information provided by the APS channel regarding the flame front is limited, and even simple subtraction of the screens yields insignificant differences. On the other hand, Fig.~\ref{fig: results2}(b) displays the comprehensive NHG obtained from the DVS channel, which records the detailed propagation behavior of the flame front. The accumulated event frame within a window of interest of 14 ms provides a rough outline of the flame front. By further reducing the window size to \SI{2.8}{ms}, the scatter plot reveals the rough structure of the flame front. A comparison between the outputs of the APS and DVS channels can be found in Video 1.

Subsequently, we compare the results in terms of the absolute velocity map and the directional pseudo-color velocity map using centroid tracking and NeuroSH methods. As depicted in Fig.~\ref{fig: results2}(c), the centroid method extracts only a few flame front structures, which are also blurred, indicating that the tracking method is only effective in the region of Type \rom{1} DR.  On the other hand, NeuroSH successfully captures the morphology of the large-gradient hemispherical flame front, where Type \rom{2} and Type \rom{3} DR dominate. Consequently, the results obtained from NeuroSH align with the flame front formation theory under laminar burning conditions. Extensive outcomes showcasing the dynamic flame front propagation behavior are included in the last section of Supplement 1 and Video 2.

\section{CONCLUSION AND DISCUSSION}
In conclusion, we develop NeuroSH, which is a physics-informed wavefront sensing framework based on the bio-inspired CNI paradigm. NeuroSH enables ultra-fast and HDR wave normal sensing, which successfully handles all three types of DR. To achieve this, NeuroSH first utilizes the proposed low-cost temporal diversity strategy based on the CNI paradigm to efficiently gain data dimensionality. Then it leverages a modified global warping scheme to extract the optimal flow that produces the clearest IWE of each sub-aperture. By approaching the 3D trajectory within the spatio-temporal neighborhood as the high-dimensional wave normal, NeuroSH can resolve the inherent ambiguity in displacement distinction. The conduction of an adaptable SH setup on a phase-only SLM provides reliable experimental verification of NeuroSH. Furthermore, NeuroSH is utilized for instantaneous flame front odometry in a BOS setup, showcasing its remarkable advantage of monitoring rapidly changing phase flows with a large DR.

In particular, we emphasize that the proposed differentiable forward model is an important innovation, which establishes a bridge between the neuromorphic modality and the SH optical configuration.  Due to the property of the utilized contrast maximization scheme, the implementation of  NeuroSH is applicable for all angle-based wavefront sensing models without specifying the mask pattern. Furthermore, the obtained motion-corrected image can facilitate the decoupling between the moving targets and the static background under photon-starved scenarios. As such, NeuroSH is widely applicable in enhancing  3D tracking and quantitative phase imaging applications, including holographic particle tracking~\cite{chen2021holographic, chen2021snapshot, huang2022recent,zhang2022holographic}, speckle analysis~\cite{ge2021event}, and non-line-of-sight imaging~\cite{metzler2020deep,zhu2023removing}.

% \appendix{}

\begin{backmatter}
	
\bmsection{Funding}
The work is supported by the Research Grants Council of Hong Kong (GRF 17201620), ACCESS --- AI Chip Center for Emerging Smart Systems, sponsored by InnoHK funding, Hong Kong SAR.

\bmsection{Acknowledgments}
We thank Haosen Liu, Zhou Ge, Dr. Hai Gong,  Prof. Evan Peng, and Prof. Wolfgang Heidrich for their technical support and useful discussions. J. H appreciates the partial financial support from Shanghai Jiao Tong University and the Hong Kong Scholars Program (XJ2022032).

\bmsection{Disclosures}
The authors declare no conflicts of interest.

\bmsection{Data Availability Statement}
Data underlying the results presented in this paper are not publicly available at this time but may be obtained from the authors upon reasonable request.  Codes will be released once published.

\bmsection{Supplemental document} See Supplements 1, Video 1, and 2 for supporting content.
 
\end{backmatter}

% \input{content/appendix}
%%%%%%%%%% If using BibTeX:
%\bibliography{zotero}
\bibliography{main}

%%%%%%%%%% If preparing manually:
% \begin{thebibliography}{1}
% \newcommand{\enquote}[1]{``#1''}

% \bibitem{Zhang:14}
% Y.~Zhang, S.~Qiao, L.~Sun, Q.~W. Shi, W.~Huang, L.~Li, and Z.~Yang,
%   \enquote{Photoinduced active terahertz metamaterials with nanostructured
%   vanadium dioxide film deposited by sol-gel method,}
%   {\protect\JournalTitle{Optics Express}} \textbf{22}, 11070--11078 (2014).

% \bibitem{Optica}
% {Optica}, \enquote{{Optica Publishing Group},}
%   \url{http://www.opg.optica.org}.

% \bibitem{FORSTER2007}
% P.~Forster, V.~Ramaswamy, P.~Artaxo, T.~Bernsten, R.~Betts, D.~Fahey,
%   J.~Haywood, J.~Lean, D.~Lowe, G.~Myhre, J.~Nganga, R.~Prinn, G.~Raga,
%   M.~Schulz, and R.~V. Dorland, \enquote{Changes in atmospheric consituents and
%   in radiative forcing,} in \enquote{Climate Change 2007: The Physical Science
%   Basis. Contribution of Working Group 1 to the Fourth assesment report of
%   Intergovernmental Panel on Climate Change,}  S.~Solomon, D.~Qin, M.~Manning,
%   Z.~Chen, M.~Marquis, K.~B. Averyt, M.~Tignor, and H.~L. Miler, eds.
%   (Cambridge University Press, 2007).

% \end{thebibliography}

\end{document}